\documentclass[conference, 10pt]{IEEEtran}

\usepackage{cite}
\usepackage[cmex10]{amsmath}
\usepackage{url}
\usepackage{graphicx}
\usepackage{color}
\usepackage{placeins}
\usepackage{float}
\usepackage{tabularx,colortbl}
\usepackage{ifthen}
\usepackage{algorithm}
\usepackage{algpseudocode}
\makeatletter

\newcounter{author}
\renewcommand{\author}[2][]{
   \stepcounter{author}
   \@namedef{author@\theauthor}{#2}
   \@namedef{authorlabel@\theauthor}{#1}
}

\newcounter{address}
\newcommand{\address}[2][]{
   \stepcounter{address}
   \@namedef{address@\theaddress}{#2}
   \@namedef{addresslabel@\theaddress}{#1}
}

\newcommand{\alsep}{and}

\def\newmaketitle{\par%
  \begingroup%
  \normalfont%
  \def\thefootnote{}
  \def\footnotemark{}
  \let\@makefnmark\relax
  \footnotesize
  \footnotesep 0.7\baselineskip
  \normalsize%
  \twocolumn[\thenewmaketitle\@IEEEaftertitletext]%
  \if@IEEEusingpubid
     \enlargethispage{-\@IEEEpubidpullup}%
  \fi
  \endgroup
  \setcounter{footnote}{0}\let\maketitle\relax\let\@maketitle\relax
  \gdef\@thanks{}%
  \let\thanks\relax}

\def\thenewmaketitle{
  \newpage
  \begin{center}%
    \vskip0.2em{\Huge\@IEEEcompsoconly{\sffamily}\@IEEEcompsocconfonly{\normalfont\normalsize\vskip 2\@IEEEnormalsizeunitybaselineskip
   \bfseries\large}\@title\par}\vskip1.0em\par%
    \vspace{1ex}
    \newcounter{c@author}
    \newcounter{c@tmp}
    \ifthenelse{\value{author}=2}{%
      \newcommand{\liand}{ and }}{%
      \newcommand{\liand}{, and }}
    \ifthenelse{\value{address}<2}{%
      \@nameuse{author@1}%
      \stepcounter{c@author}%
      \whiledo{\value{c@author}<\value{author}}{%
        \setcounter{c@tmp}{\value{author}}%
        \addtocounter{c@tmp}{-\value{c@author}}%
        \ifthenelse{\value{c@tmp}=1}{%
          \renewcommand{\alsep}{\liand}}{\renewcommand{\alsep}{, }}%
        \stepcounter{c@author}\alsep \@nameuse{author@\thec@author}}\\%
    }
    {
      \@nameuse{author@1}${}^{(\ref{\@nameuse{authorlabel@1}})}$%
      \stepcounter{c@author}%
      \whiledo{\value{c@author}<\value{author}}{%
      \setcounter{c@tmp}{\value{author}}%
      \addtocounter{c@tmp}{-\value{c@author}}%
      \ifthenelse{\value{c@tmp}=1}{%
        \renewcommand{\alsep}{\liand}}{\renewcommand{\alsep}{, }}%
      \stepcounter{c@author}\alsep \@nameuse{author@\thec@author}%
        ${}^{(\ref{\@nameuse{authorlabel@\thec@author}})}$%
      }
    }
    \vspace{0.2ex}

    \ifthenelse{\value{address}>0}{%
      \ifthenelse{\value{address}=1}{
        {\@nameuse{address@1}}
      }
      {
        \newcounter{c@address}

        \begin{center}
        \whiledo{\value{c@address}<\value{address}}
        {
          \refstepcounter{c@address}
            ${}^{(\thec@address)}$\,%
              \label{\@nameuse{addresslabel@\thec@address}}%
              \@nameuse{address@\thec@address}\\ %
        }
        \end{center}
      } 
    }
    {
      \relax
    }
  \end{center}
}

\makeatother

\title{Exploring Radial Symmetry on Phased Arrays Using Particle Swarm Optimization}

\author[org1]{Arkadii KAZANSKII}
\author[org1]{Juan Andres VASQUEZ-PERALVO}
\author[org1]{Symeon CHATZINOTAS}
\address[org1]{SigCom, SnT, University of Luxembourg, Luxembourg,
(arkadii.kazanskii, juan.vasquez, symeon.chatzinotas)@uni.lu}

\begin{document}

\newmaketitle

\begin{abstract}
Phased antenna arrays enable dynamic beam shaping, which is essential for Non-Geostationary (NGSO) satellite communications where efficient beam distribution is important. This study focuses on thinning phased antenna arrays with circular apertures made up of eight replicated sectors. Circular apertures reduce the number of active elements, lowering system costs and improving radiation performance by evenly distributing energy, which helps to reduce Side Lobe Levels (SLL).

Particle Swarm Optimization was used to approach the thinning problem, addressing the challenge of selecting which elements should be activated. The resulting design achieves an SLL of \(-25.67 \, \mathrm{dB}\), outperforming previous designs with SLL reductions of \(-22.53 \, \mathrm{dB}\).

Achieved results underscore the potential of circular aperture phased arrays to improve beam quality, minimize interference, and deliver cost-effective solutions for NGSO satellites.
\end{abstract}

\section{Introduction}

To maintain the quality of modern satellite communication services, dynamic steering and beamwidth control of the radiation pattern are essential. This capability is particularly critical in NGSO communications, where the relative motion of the satellite with respect to the coverage area requires continuous beam spot adaptation \cite{10491591},\cite{duncan2023harnessing}, \cite{ha2024user}. 

To achieve the desired beaming capabilities while minimizing complexity, costs, weight, computational demands, and antenna size — factors that are particularly critical in Low Earth Orbit (LEO) satellite systems — thinning of phased planar antenna arrays offers an effective solution \cite{10491591}, \cite{10197375}, \cite{mailloux2017phased}.

In current work, we're focusing on the design of a phased antenna array with a circular aperture made of 8 replicated \(45^\circ\) sectors of antenna elements with a rectangular distribution. 

Circular aperture antenna arrays require fewer elements than a similar one with a square aperture, thus reducing its cost while enhancing the radiating characteristics by reducing Side Lobe Level (SLL) through equal energy distribution across the aperture \cite{9425132}, \cite{mailloux2017phased}, \cite{10197375}. 

\begin{figure}[hbt!]
    \centering    \includegraphics[width=\linewidth]{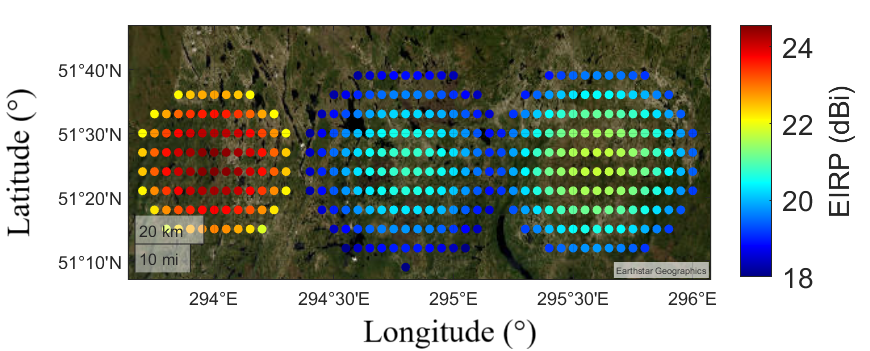}
    \caption{Beam Spot Allocation Example}
    \label{fig:beams}
    \vspace{-15pt}
\end{figure}

Radial symmetry of such arrays provides an opportunity for utilization in efficient beaming with interference reduction through placing nulls in the main lobes of the neighboring spots as illustrated in Fig. \ref{fig:beams}.
\vspace{-0cm}

The task of antenna array thinning involves finding a combination of switched "On" and "Off" elements of the antenna array to achieve a desired radiation pattern. This task is non-convex and could be solved using different algorithms, such as Genetic Algorithm (GA), Particle Swarm Optimisation (PSO), etc. A work by Panduro et al. \cite{panduro2009comparison} notes that PSO (applied to the design of circular antenna arrays) achieves better results than other methods.
PSO, initially introduced by Kennedy and Eberhart~\cite{kennedy1995particle}, a bio-inspired algorithm, is used in this study for its simplicity and robustness in exploring large solution spaces. 

Several studies have explored phased array thinning using various algorithms \cite{gal2018thinning}, \cite{mandal2012thinned}, achieving sidelobe levels (SLL) as low as \(-22.53\ \text{dB}\) in cases where axi-symmetric pattern optimization was maintained.

In more recent works \cite{LI2023120771} and \cite{sun2023synthesis}, SLL values ranging from \(-24.10\ \text{dB}\) to \(-33.62\ \text{dB}\) were reported. However, the resulting radiation patterns in these studies lacked the desired axi-symmetric characteristics.

In the present work, an SLL reduction down to \(-25.67\ \text{dB}\) was achieved for a configuration consisting of 673 antenna elements, while preserving the axi-symmetric pattern requirements.

\section{Approach}\label{sec:approach}

\begin{figure*}[hbt!]
    \centering    \includegraphics[width=0.95\textwidth]{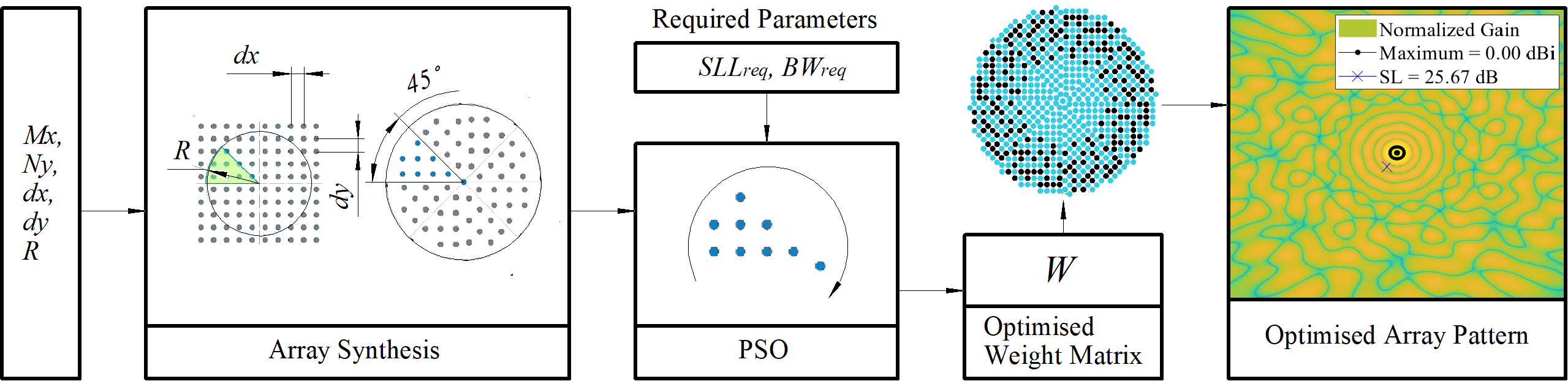}
    \caption{Research Pipeline}
    \label{fig:prob_desc}
    \vspace{-10pt}
\end{figure*}

Fig. \ref{fig:prob_desc} showcases the overall pipeline of the process used in this research. 

The design begins with a set of array parameters, such as element spacing (\(dx, dy\)), the number of elements in the x and y directions (\(M_x, N_y\)), operating frequency, and aperture radius (\(R\)). These parameters are used to synthesize a circular aperture antenna array by initially creating a 45-degree sector of elements and then replicating it seven times around the origin to form a full circular array as described in subsection \ref{sec:array_synth}.

The array synthesis is then followed by a thinning algorithm, implemented using the Particle Swarm Optimization (PSO) technique, which is described in subsection \ref{sec:PSO}. The algorithm optimizes the weight matrix (\(W\)) based on input parameters such as the desired sidelobe level (\(SLL_{\text{req}}\)) and beamwidth (\(BW_{\text{req}}\)). Activation of the elements in the initial sector of 45\(^o\) is iterated.

The optimized weight matrix results in an efficient radiation pattern, as shown in the section \ref{sec:res}.

This section details the approach employed to address the stated problem, including the antenna array design process, and the research framework for optimization and analysis in MATLAB.

\subsection{Antenna Array Synthesis}\label{sec:array_synth}

The synthesis process for a generic circular aperture antenna array of a radius \(R\) consisting of 45-degree sectors with a central element follows the next steps:

\begin{enumerate}
\item \textbf{Creating a 45-degree array sector}. It starts with the creation of a uniform rectangular array, out of which a 45-degree sector of a certain radius \(R\) was cut. The elements that lie inside the sector and on its borders were preserved (marked green in Fig. \ref{fig:array_design}, left).  

\begin{figure}[hbt!]
    \centering
    \includegraphics[width=\linewidth]{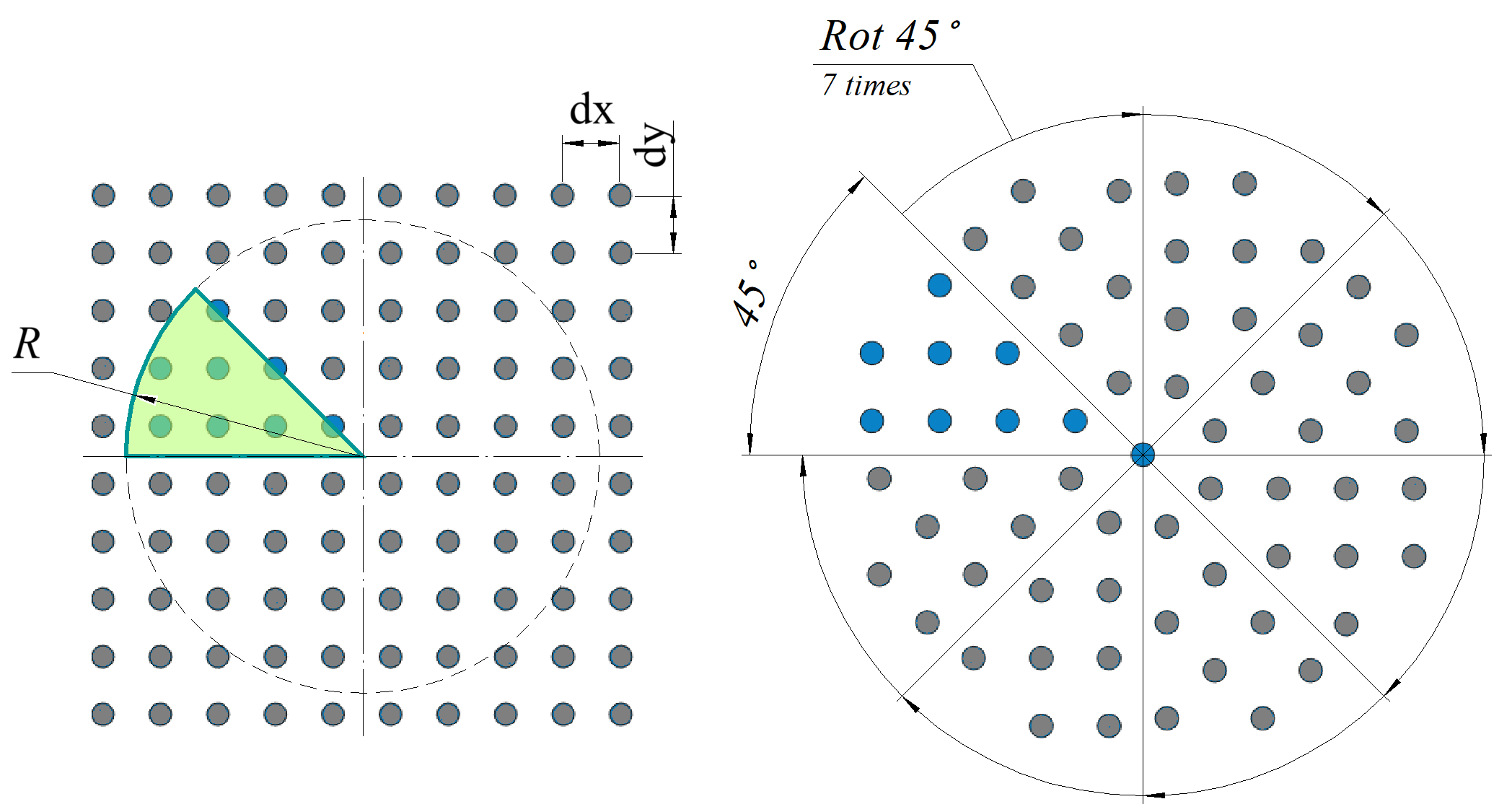}
    \caption{Circular aperture array antenna design example, 45-degree sector cutout (left), full aperture - 65 resulting elements (right)}
    \label{fig:array_design}
    \vspace{-10pt}
\end{figure}

\item \textbf{Creating a full circular aperture}. This sector was then shifted and replicated seven times around the central axis by applying successive 45-degree rotational transformations, resulting in a complete circular array. Additionally, a central element was added to the array in the origin (as portrayed in Fig. \ref{fig:array_design}, right). 
\end{enumerate}

\subsection{Research Framework}\label{sec:res_frw}

A variety of antenna array configurations, characterized by different radii, were analyzed using the proposed pipeline as illustrated in Figure \ref{fig:prob_desc}. The analysis provided optimized weight matrices (\(W\)), along with the corresponding side lobe level (SLL) and beamwidth (BW) values for each configuration, as seen in section \ref{sec:res}.


\section{Problem Formulation}\label{sec:prob_desc}

Current section outlines the mathematical formulation of the planar array radiation pattern, as well as the optimization approach employed to achieve a thinned array configuration.
 
\subsection{Planar Array Radiation Pattern}

Consider a 2D antenna array with \(N_y*M_x\) elements, each spaced by period \(d\). Let \(\lambda_0\) be the wavelength at the carrier frequency \(f\). Inter-element spacing of \(d = \lambda_0/2\) was implied in order to avoid the onset of grating lobes and decrease mutual coupling. Therefore, the total radiation pattern \(AP(\theta, \phi)\) of this planar antenna array can be computed using the array pattern formula described in Eq. \ref{eq:AP} \cite{mailloux2017phased}:

\begin{equation}
AP(\theta, \phi) 
= E_{UC}(\theta, \phi) \times AF(\theta, \phi),  
\label{eq:AP}
\end{equation}

where \(E_{UC}(\theta, \phi)\) represents the element pattern, modeled in this research as a cosine element and calculated as shown in Eq. \ref{eq:UC}, and \(AF(\theta, \phi)\) denotes the array factor, computed as described in Eq. \ref{eq:AF}.

\begin{equation}
E_{UC}(\theta, \phi) 
= \cos^2\theta\cos^2\phi,
\label{eq:UC}
\end{equation}

\begin{equation}
\begin{split} 
AF(\theta, \phi) 
= \sum_{m=0}^{M_x}\sum_{n=0}^{N_y} w_{mn} \, e^{\,j(m)(k d_x (\sin\theta \cos\phi))}\times \\ \times e^{\,j(n)(k d_y (\sin\theta \cos\phi))},   
\end{split}
\label{eq:AF}
\end{equation}

where:
\begin{itemize}
    \item \(M_x\) and \(N_y\) are numbers of elements in x and y directions,
    \item \(d_x\) and \(d_y\) are the periods of the elements in x and y directions,
    \item \(w_{mn}\) is the complex weight of the \(mn\)-th element,
    \item \(k = \frac{2\pi}{\lambda_0}\) is the wavenumber,
    \item \(\theta\) and \(\phi\) are elevation and azimuth angles in spherical coordinates.
\end{itemize}

For the \emph{thinning} problem, \(w_{mn}\in \{0,1\}\). In order to model a circular aperture, the weights of the elements \(m,n\) outside of the circular aperture (as described in \ref{sec:approach}) are set to be \(w_{mn} = 0\).

\subsection{Taylor Taper as a Benchmark}

The Taylor taper is commonly used to reduce sidelobes while controlling the main-lobe width. For a 1D array of length \(N\), the Taylor weight vector \(T(n)\) is  expressed as in Eq. \ref{eq:taylor}:
\begin{equation}
T(n) = \frac{I_0 \!\Bigl( \alpha \sqrt{1 - \Bigl(\frac{2n}{N}\Bigr)^2} \Bigr)}
             {I_0(\alpha)},
\quad n = -\frac{N}{2}, \ldots, \frac{N}{2}-1, 
\label{eq:taylor}
\end{equation}
where \(\alpha\) is often chosen based on the desired sidelobe level (For a Taylor taper with -30 dB sidelobes, \(\alpha\) is typically set around 3-4), and \(I_0\) is the modified 0-order Bessel function of the first kind. 

\subsection{Particle Swarm Optimization Application}\label{sec:PSO}

Algorithm \ref{alg:pso} is applied for the thinning problem :

\begin{algorithm}
\caption{Particle Swarm Algorithm applied to Antenna Array Thinning}\label{alg:cap}
\begin{algorithmic}[1]
\item[\textbf{Input}] $SLL_{req}, BW_{req}, threshold$
\item[\textbf{Output}] $W$\Comment{weight matrix, based on the best particle}; $\textbf{X}_p,\textbf{V}_p,SLL_p, BW_p$
\State \textbf{Initiate} $\textbf{X}_p(0),\textbf{V}_p(0),$
\State ${X}_p \gets {X}_p(0)$
\State ${V}_p \gets {V}_p(0)$
\State $t \gets 0$
\While{$F_p(SLL_p(t),BW_p(t)) (Eq. \ref{eq:costfunction}) \geq threshold$}
    \State \textbf{Calculate:} Antenna Pattern $AP_p(\theta = 0, \phi, t)$ for each particle 
    \State \textbf{Extract:} $SLL_p(t),BW_p(t)$
    \State \textbf{Update:} ${V}_p(t+1)$  (Eq. \ref{eq:Vp})
    \State \textbf{Update:} ${X}_p(t+1)$  (Eq. \ref{eq:Xp+1})
    \State \textbf{Update:} $t = t+1$
\EndWhile
\State Get Optimised W from ${X}_p(t)$
\end{algorithmic}
\label{alg:pso}
\end{algorithm}

\subsubsection{Particle Representation}

Each "particle" in the swarm is a \(\tfrac{N}{8}\) vector of binary values. The position vector of a single particle is denoted in Eq. \ref{eq:Xp}:
\begin{equation}
\mathbf{X}_p = \bigl[x_{p,1}, x_{p,2}, \ldots, x_{p,D}\bigr]^\top \in \{0,1\}^D,
\label{eq:Xp}
\end{equation}
where \(D = \tfrac{N-1}{8} + 1 \) is the dimensionality (the total number of elements in the sector and the central element). 

\subsubsection{Velocity Update}

The particle velocity \(\mathbf{V}_p\) at iteration \(t\) is updated according to the canonical PSO rule (Eq. \ref{eq:Vp}) \cite{kennedy1995particle}:
\begin{equation}
\begin{split}
\mathbf{V}_p(t+1) 
= \omega \,\mathbf{V}_p(t)
+ c_1\,r_1\bigl(\mathbf{P}_p - \mathbf{X}_p(t)\bigr) +\\
+ c_2\,r_2\bigl(\mathbf{G} - \mathbf{X}_p(t)\bigr),    
\end{split}
\label{eq:Vp}
\end{equation}
where:
\begin{itemize}
    \item \(\omega\) is the inertia weight (set as \textit{0.75} \cite{chen2005synthesis}),
    \item \(c_1, c_2\) are the cognitive and social learning factors (both set at \textit{2.5} \cite{chen2005synthesis}),
    \item \(r_1, r_2 \in [0,1]\) are random scalars,
    \item \(\mathbf{P}_p\) is the personal best position of particle \(p\) (\(p | \max(F(X_p))\)),
    \item \(\mathbf{G}\) is the global best position found by the entire swarm (\(p | \max(F)\)).
\end{itemize}

\subsubsection{Position Update}

The next position of the particle is calculated as given in Eq. \ref{eq:Xp+1} \cite{kennedy1995particle}:
\begin{equation}
X_p(t+1)= X_p(t) + \chi V_p(t+1),
\label{eq:Xp+1}
\end{equation}
where \(\chi\) is the constriction coefficient. In our implementation, \(\chi = 0.75\)\cite{chen2005synthesis}. 

An additional step done was the binarisation of the values of vector \(X_p(t+1)\), due to the use of classical PSO.

\subsubsection{Fitting Function}
 
A fitting function for each particle \(F_p\) for each time step t is formulated as in Eq. \ref{eq:costfunction}:
\begin{equation}
\begin{split}
F_p(t) 
= w_1( \mathrm{SLL}(\mathbf{X}_p) - \mathrm{SLL}_\text{req}\bigr)^2 + \\  +  w_2 \bigl(\mathrm{BW}(\mathbf{X}_p) - \mathrm{BW}_\text{req} \bigr)^2,    
\end{split}
\label{eq:costfunction}
\end{equation}
where \(w_1, w_2\) are relative weights for each criterion. 
\section{Results and Discussion}\label{sec:res}

This section presents the results for arrays with radii ranging from \(R = 4\lambda_0\) to \(R = 15\lambda_0\), corresponding to 41 to 673 elements. The inter-element spacings are \(d_x = d_y = \lambda_0/2\), and the operating frequency is set to 12 GHz (Ku band).

\begin{figure}[hbt!]
    \centering
    \includegraphics[width=0.5\textwidth, trim={0 0 0 34pt}, clip]{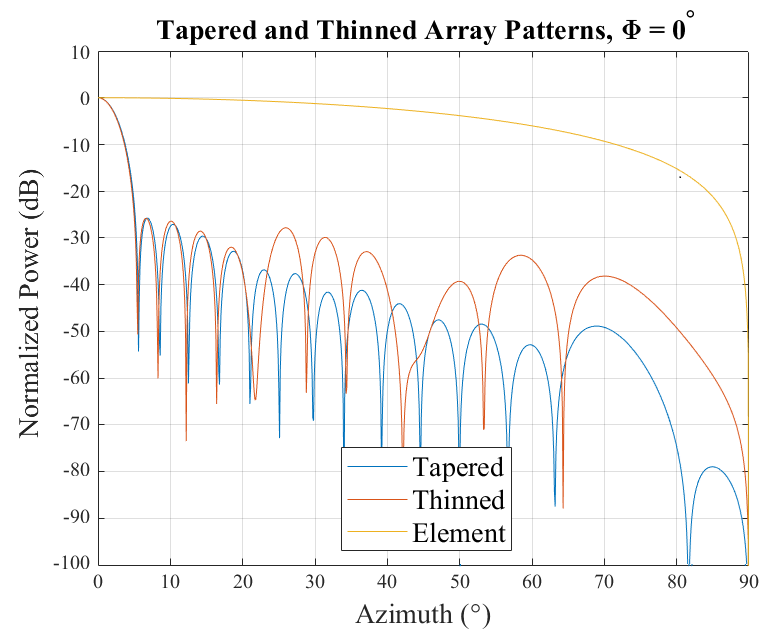}
    \caption{Tapered and Thinned Array Pattern Comparison, \(\Phi = 0^{\circ}, \Theta = 0^{\circ}\)}
    \label{fig:TaperedvsOptimisedAz}
    \vspace{-10pt}
\end{figure}

Fig. \ref{fig:TaperedvsOptimisedAz} shows a comparison plot between the \emph{Thinned} and \emph{Tapered} array patterns at \(\Phi = 0^\circ, \Theta = 0^\circ\) for an array populated by \(N = 673\) elements. For comparison, a phased array utilizing amplitude tapering is used, as this technique effectively controls the side-lobe levels (SLL) by adjusting the excitation amplitude of each element. 

The angular cuts at \(\Phi = 0^\circ, \Theta = 0^\circ\) are selected due to the symmetrical properties of these arrays, as these directions are where the maximum SLL is expected to occur.

The PSO-optimized array pattern closely matches the main lobe of the tapered array, effectively maintaining the desired pattern. However, the sidelobes are slightly elevated and display increased irregularity across angular regions. These deviations can be attributed to interference effects caused by the binary activation of elements and mutual coupling between elements, which distort the smooth transitions observed in the tapered design.

\begin{figure}[hbt!]
\centering
\includegraphics[width=0.35\textwidth]{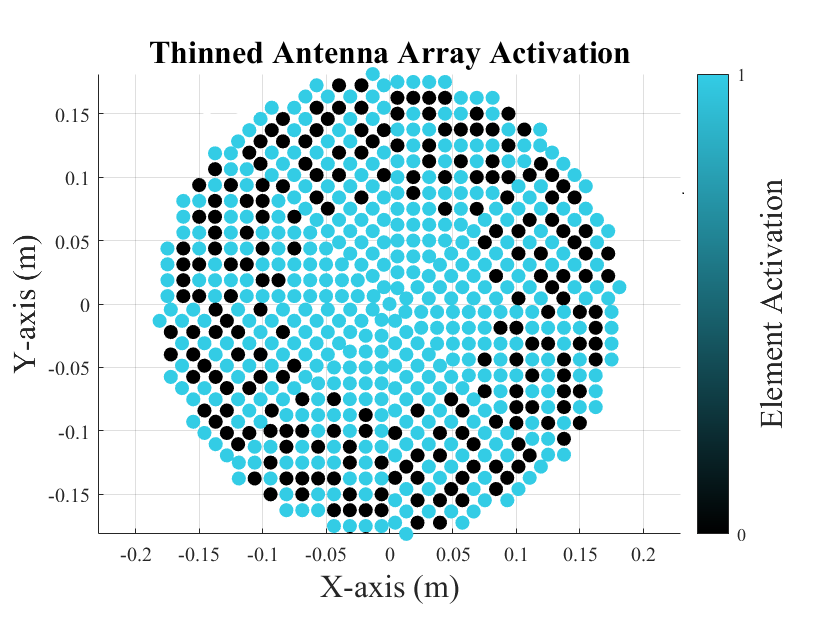}
\caption{Thinned antenna array activation, 673 elements. 457 elements are activated.}
\label{fig:activation_tap}
\vspace{-10pt}
\end{figure}

\begin{figure}[hbt!]
\centering
\includegraphics[width=0.35\textwidth]{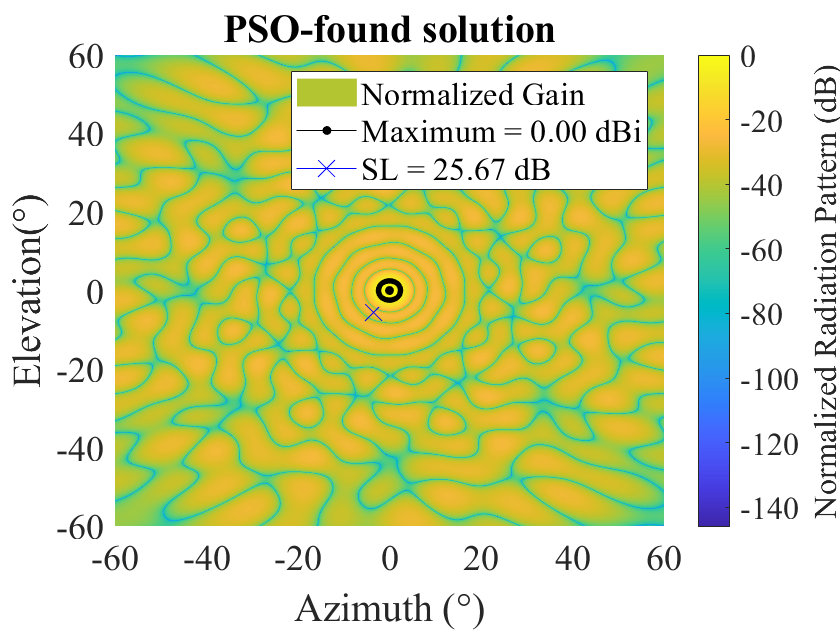}
\caption{3D Radiation Pattern of the thinned array, 673 elements. 457 elements are activated.}
\label{fig:3d_thinned}
\vspace{-10pt}
\end{figure}

Fig. \ref{fig:activation_tap}, \ref{fig:3d_thinned} show the activation and the 3D radiation pattern of the above-mentioned thinned array.

\begin{figure}[hbt!]
    \centering    \includegraphics[width=0.45\textwidth]{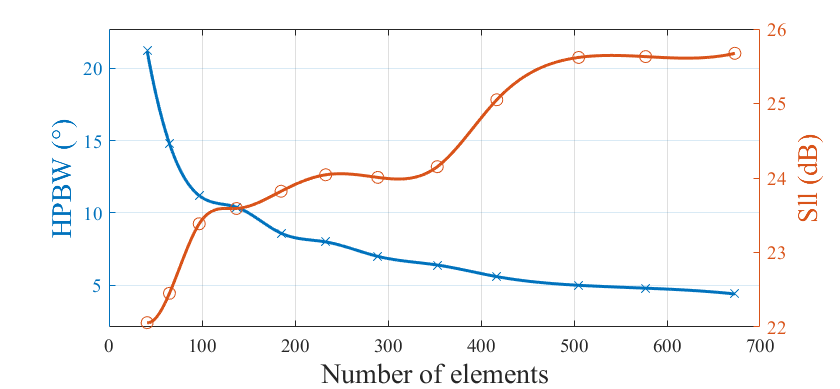}
    \caption{Relation between HPBW, SLL and the number of elements in an array}
    \label{fig:SLL_BWvsN}
    \vspace{-20pt}
\end{figure}

One of the key findings of this study is the relationship between the achieved Side Lobe Levels (SLL) in thinning via Particle Swarm Optimization (PSO), the Half-Power Beamwidth (HPBW), and the number of antenna elements in the array (as illustrated in Fig. \ref{fig:SLL_BWvsN}). The results indicate that the SLL does not continuously rise with the increasing number of elements; instead, it stabilizes at specific array sizes. This behavior may relate to the non-convex nature of the thinning problem, which can lead the optimizer to saddle points rather than the global minimum.

\section{Conclusion}
This study highlights the effectiveness of Particle Swarm Optimization (PSO) in thinning circular aperture antenna arrays, achieving significant sidelobe level (SLL) reductions while preserving beam symmetry.

The achieved features demonstrate the potential of circular aperture phased arrays to enhance beam quality, reduce interference, and provide cost-effective solutions for NGSO satellite systems.

\section*{ACKNOWLEDGEMENT}
This work was supported by the Luxembourg National Research Fund (FNR), through the CORE Project ($C^3$): Cosmic Communication Construction under Grant C23/IS/18116142.



\bibliographystyle{IEEEtran}
%

\bibliography{bibliography}


\end{document}